\documentclass[twocolumn,showpacs,preprintnumbers,amsmath,amssymb]{revtex4}

\newcommand{\Hi}{\mathcal{H}}

\newcommand{\lv}{\left \vert}
\newcommand{\rv}{\right \vert}
\newcommand{\la}{\left \langle}
\newcommand{\ra}{\right \rangle}
\newcommand{\ket}[1]{\lv #1 \ra}
\newcommand{\bra}[1]{\la #1 \rv}

\begin{document}
\title{Entanglement convertibility for infinite dimensional pure bipartite states}
\author{Masaki Owari$^{1,3}$, Keiji Matsumoto$^{2,3}$ and Mio
Murao$^{1,4}$}
\address{
$^1${\it Department of Physics, The University of Tokyo, Tokyo 113-0033, Japan}\\
$^2${\it Quantum Computation Group, The National Institute of Information, Tokyo 101-8430, Japan}\\
$^3${\it Imai Quantum Computation and Information Project, JST, Tokyo 113-0033, Japan}\\
$^4${\it PRESTO, JST, Kawaguchi, Saitama 332-0012, Japan} }
\date{\today}

\begin{abstract}
It is shown that the order property of pure bipartite states under
SLOCC (stochastic local operations and classical communications)
changes radically when dimensionality shifts from finite to
infinite. In contrast to finite dimensional systems where there is
no pure incomparable state, the existence of infinitely many
mutually SLOCC incomparable states is shown for infinite
dimensional systems even under the bounded energy and finite
information exchange condition. These results show that the effect
of the infinite dimensionality of Hilbert space, the ``infinite
workspace'' property, remains even in physically relevant infinite
dimensional systems.
\end{abstract}


\maketitle

%%%%%%%%%%%%introduction%%%%%%%%%%%%%%%%%%%%%%%%%%%%%%%%%%%%%%%%

Quantum information encoded in quantum systems offers possibility
to provide a new and outstanding class of the information
processing. Quantum two-level systems (qubit systems), which
correspond to the binary `bit' systems of classical information
processing, are often used for quantum information processing, but
quantum information processing can also be performed in finite
multi-level systems and even infinite dimensional systems, such as
bosonic systems.

Infinite dimensional systems, sometimes called continuous variable
systems, have been expected to offer high potential for quantum
information processing. One of the advantages of infinite
dimensional systems is the possibility of implementation using
quantum optical systems, as shown by the successful demonstration
of teleportation \cite{teleportation}. Properties of quantum
information under gaussian operations, which can be implemented by
linear optical systems, have been investigated \cite{gaussian,
gaussian distill}. Another advantage is in its infinite
dimensionality of Hilbert space. Since finite multi-level systems
can be reduced to (finitely) many two-level systems, the essential
power of quantum information processing is not different from
qubit systems, although there may be advantages of tractability in
implementation.  However, infinite dimensional systems cannot be
reduced to two-level systems in general. Therefore, there is the
possibility of (yet unknown) new types of quantum information
processing schemes, which do not exist in finite dimensional
systems.

It has been widely believed that the fundamental properties of
finite and infinite dimensional systems are similar or, at least,
that in physically relevant infinite dimensional systems with
bounded energy conditions and with a finite number of measurement
outcomes, the properties can be well approximated to finite
dimensional systems. In this letter, we will show that these
beliefs are not well-founded for quantum systems by investigating
entanglement properties of infinite dimensional systems.

Entanglement is regarded as the key resource which allows many
quantum information processing schemes out-perform than their
classical counterparts. Entanglement can be classified by
transformability  under local operations, such as LOCC
(deterministic local operations and classical communications),
SLOCC (stochastic LOCC). The convertibility properties of two
different entangled states (in a single copy or multi-copy
situation) under local operations are important for the
qualitative and quantitative understanding of entanglement. For
finite dimensional bipartite systems, we now have a better
understanding of LOCC and SLOCC convertibility based on intensive
work in recent years.  For example, the condition for the
convertibility of two pure entangled states in the single copy
situation is given by Nielsen's majorization theorem
\cite{nielsen} for LOCC, and is given by Vidal's theorem
\cite{vidal} for SLOCC.  It is also found that incomparable pure
states only exist in multipartite systems, such as the GHZ state
$(\ket{000}+\ket{111})/\sqrt{2}$ and the W state
$(\ket{001}+\ket{010}+\ket{100})/\sqrt{3}$ in three qubit systems
\cite{dur}.

In this letter, we will show that a fundamental property of
entanglement, the order properties of pure bipartite states under
SLOCC are changed significantly from total order to non-total
order with the shift in dimensionality from finite to infinite.
Further, we will show that there are infinitely many mutually
SLOCC incomparable pure bipartite states even under the bounded
energy condition. These results show that the effect of infinite
dimensionality of Hilbert space, the ``infinite workspace''
property, remains in physically relevant infinite dimensional
systems.

%%%%%%%%convertibility and set theory%%%%%%%%%%%%%%%%%%%%%%%%%

First, we consider ordering sets, ordered by the convertibility of
two general bipartite quantum states under general operations.  In
the language of set theory \cite{settheory}, convertibility can be
describe by an order denoted by ``$\rightarrow$''. For a set of
physical states $S$, the order indicating the existence of
physical transformation satisfies the reflective law $a
\rightarrow a$ and the transitive law $a \rightarrow b$ and $b
\rightarrow c$ imply $a \rightarrow c$), where $a, b, c \in S$.
This ordering property is a pseudo partial ordering. We consider
that two states are in the same equivalence class, if they
transform each other $a \rightarrow b$ and $b \rightarrow a$. We
denote this situation as $a \leftrightarrow b$. The quotient set
of $S$ by the equivalence class $\leftrightarrow$ is denoted as
$(S/\leftrightarrow , \rightarrow)$ and represents the
classification based on the given transformation.  For the set
$(S/\leftrightarrow , \rightarrow)$, we can redefine the ordering
$\rightarrow$ which satisfies the additional condition $a
\rightarrow b$ and $b \rightarrow a$ imply $a = b$. This ordering
property is partial ordering.

If the set has the additional property that $a \nrightarrow b$
implies $b \rightarrow a$ ($a \nrightarrow b$ denotes $a
\rightarrow b$ is not true), the set is totally ordered.  Total
ordering is a convenient property to analyze the convertibility of
a system, since there exists a unique measure of ordering for a
totally ordered set. For example, the convertibility of pure state
$\ket{\phi}$ under LOCC transformation in the asymptotic situation
is total ordering. Therefore, there is a unique measure, or a
monotone, given by the von Neumann entropy of entanglement
$E(\ket{\phi})$ in this case \cite{thermodynamics}. On the other
hand, a pair of monotones, like distillable entanglement and
entanglement cost \cite{bennett}, are useful tools to distinguish
the ordering properties (total ordering or partial ordering) of
the system \cite{morikoshi}.

%%%%%%%%%%%%%%%%%%%ordering-monotone%%%%%%%%%%%%%%%%%%%%%%%%%%%%%%%%

The key idea of our formulation is that we consider a totally
ordered subset $\{ \xi _r \}$ parameterized by a real number $r$
for a pseudo partial ordering set $S$.  Then we can always define
a pair of functions $R^-(\psi)$ and $R^+(\psi)$ for a state $\psi
\in S$, where $R^-(\psi)$ is the supremum of $r$ at which $\psi$
can be transformed to $\xi _r$, and $R^+(\psi)$ is the infimum of
$r$ at which $\Psi$ can be transformed to $\xi _r$. Mathematically
they are expressed as the following: For a pseudo partial ordering
set, if there exists a real parameterized total ordering subset
$\{ \xi _r \} _{r \in A} \subset S$ where $A \subset \mathbb{R}$
such that $r_1 \le r_2$ if and only if $\xi _{r_1} \rightarrow \xi
_{r_2}$, we can define a pair of functions on $S$ to $\overline{A}
\subset \mathbb{R}$ as
\begin{eqnarray}
R^-(\psi) &=& \sup \{ r \in A | \psi \rightarrow \xi _r \}
\label{rminus}
\\
R^+(\psi) &=& \inf \{ r \in A | \xi _r \rightarrow \psi \}
\label{rplus}
\end{eqnarray}
where we define $R^-(\ket{\psi}) = \inf \{ A \}$ for $\{ r\ \in A
| \psi \rightarrow \xi _r \} =\emptyset $, and $R^+(\ket{\psi}) =
\sup \{ A \}$ for $\{ r\ \in A | \xi _r \rightarrow \psi \}
=\emptyset $.

Although there are many ways to define a monotone for a partial
ordering set from a totally ordered subset $\{ \xi _r \} _A$, our
definition of functions $R^-(\psi)$ and $R^+(\psi)$ are preferable
for analyzing entanglement convertibility. It is easily proven by
contradiction that they are the {\it unique monotones} which give
lower and upper bounds of any monotones defined from $\{ \xi _r
\}$ for a given pseudo partial ordered set $S$, that is, $\psi
\rightarrow \phi$ implies $R^-(\psi) \ge R^-(\phi)$ and $R^+(\psi)
\ge R^+(\phi)$, $R^+(\phi) < R^-(\psi)$ implies $\psi \rightarrow
\phi$, and also any other monotone $R_0 (\psi)$ defined from $\{
\xi _r \}$ satisfies $R^- (\psi) \le R_0 (\psi) \le R^+ (\psi)$
for all $\psi \in S$.

From the properties of $R^-(\psi)$ and $R^+(\psi)$, we can
immediately derive the following important results:  If the
quotient set $(S/\leftrightarrow , \rightarrow)$ is totally
ordered, namely, $\psi \nrightarrow \phi$ implies $\phi
\rightarrow \psi$ is satisfied, then for all $\psi \in S$, we have
$R^-(\psi) = R^+(\psi)$. On the other hand, if there exists $\psi
\in S$ such that $R^-(\psi) < R^+(\psi)$, then $(S/\leftrightarrow
, \rightarrow)$ is not totally ordered, and $\psi$ is incomparable
to all $\xi _r$ with $R^-(\psi) < r < R^+(\psi)$.

%%%%%%%%%%%%%%%%%%%%%%%%%%%%%%%%%%%%%%%%%%%%%%%%%%%%%%%%%%%%%%%%%%%%%%

The advantage of our formulation is that it can be applied for
many different situations: single-copy or asymptotic (infinitely
many-copy) cases, for finite or infinite dimensional systems,
mixed or pure states, and under LOCC, SLOCC or PPT \cite{ppt}
operations. Many important known results of entanglement theory
can be re-derived only from simple ordering properties and the
existence of the real parameterized total ordering subset. For
example, in the case of convertibility under LOCC operations for
mixed states in the asymptotic situation, we consider $\{ \ket{\xi
_s} \bra{\xi _s} \} _{s=0}^{\infty}$ to be a subset of pure states
with $E(\ket{\xi _s}) =s$ \cite{bennett} where $E(\ket{\psi})$ is
the amount of entanglement for pure states. Then we have
$R^-(\rho)= E_d(\rho)$ and $R^+(\rho)=E_c(\rho)$ where $E_d(\rho)$
and $E_c(\rho)$ are distillable entanglement and entanglement
cost, respectively. We see that the sets are not totally ordered
and $R^-$ (distillable entanglement) and $R^+$ (entanglement cost)
are limits of other monotones \cite{uniqueness}, and also that
there is a set of states such that $R^-=0$ but $R^+> 0$, the bound
entangled states \cite{boundentangle}.

%%%%%%%%%%%%%%%%%%%% slocc-monotone%%%%%%%%%%%%%%%%%%%%%%%%%%%%%%%%%%%%

Now we concentrate on the investigation of SLOCC convertibility
(with non-zero probability) of infinite dimensional pure states in
the single-copy situation.  In general, the entanglement of a pure
state $\ket{a}$ is characterized by the sequence of Schmidt
coefficients $\{ \lambda^a_I \}$ ($0 \leq I \leq d$ and $0 \leq I
\leq \infty$ for finite $d$ and infinite dimensional systems,
respectively).  For finite dimensional systems, the two monotones
coincide with the Schmidt rank, the number of non-zero Schmidt
coefficients (Vidal's theorem \cite{vidal}). On the other hand,
the Schmidt rank itself cannot be a monotone for infinite
dimensional systems for analyzing convertibility between
``genuine'' infinite dimensional states (with infinite Schmidt
ranks).  Further, if we consider exact conversion of genuine
infinite dimensional states under SLOCC, we generally need
infinite many information exchanges, which is not physical.

To extend Vidal's theorem to physical infinite dimensional
systems, we define SLOCC convertibility in infinite dimensional
systems as the following: $\ket{\psi}$ is convertible to
$\ket{\phi}$ by SLOCC if and only if $\ket{\psi}$ is convertible
to any neighborhood of $\ket{\psi}$ by finite steps of LOCC with
probability larger than some positive number.  Then Vidal's
theorem can be extended to infinite dimensional systems as the
following: $\ket{\psi} \in \Hi$ can be converted to $\ket{\phi}
\in \Hi$ by SLOCC with non-zero probability in the single-copy
situation if and only if there exists $\epsilon > 0$, $g_\psi (n)
/ g_\phi (n) \ge \epsilon$ for all $n \in N$, where $g_a (n) =
\sum _{I=n}^{\infty} \lambda^a_I$ is a function defined in terms
of Schmidt coefficients $\{ \lambda^a_I \}_{I=0}^\infty$ of a
genuine infinite dimensional state $\ket{a}$ {\bf (Vidal's
Theorem)}.  Mathematically rigorous arguments for the extended
Vidal's theorem are given in \cite{owari-locc} together with an
extension of Nielsen's theorem to infinite dimensional systems.

The function $g_a (n)$ plays the central role in the construction
of the monotones $R^-$ and $R^+$ for infinite dimensional states.
By definition, a sequence of the function $\{g_a (n)\}_{n \in N}$
satisfies four conditions, strict positivity $g_a (n)>0$, strict
monotonicity $g_a(n) > g_a(n+1)$, convexity $g_a (n+1) \le
\{g_a(n) + g_a (n+2)\}/2$, and normalization $g_a (0)=1$.
Conversely, for a given sequence of functions $\{ g(n)
\}_{n=0}^{\infty}$, there exist a genuine infinite dimensional
state $\ket{a}$, where the Schmidt coefficients are give by
$\lambda^a_I =g(n) -g(n+1)$ if and only if $\{g(n)\}_{n \in N}$
satisfies the strict positivity, strict monotonicity, convexity
and normalization conditions.  Such a function $g(n)$ is called
Vidal's monotone.

According to Vidal's theorem, if a real parameterized subset $\{
\ket{\xi _r} \} _{r \in A } \subset \Hi$ is totally ordered,
$g_{\xi_r}(n)$ must satisfies $\underline{\lim} _{n \rightarrow
\infty} (g_{\xi_{r_1}} (n) / g_{\xi_{r_2}}(n)) > 0$ if and only if
$r_1 \le r_2$ for all $r_1$ and $r_2$. From the property of
$g_{\xi_r}(n)$, we can construct the monotones $R^-$ and $R^+$
\begin{eqnarray}
    R^-(\psi) &=& \inf \{ r \in A | \underline{\lim} _{n
    \rightarrow \infty}
    g_\psi  (n)/ g_{\xi_r}(n) = 0 \}
    \label{R^-} \\
    R^+(\psi) &=& \inf \{ r \in A | \overline{\lim} _{n
    \rightarrow \infty}
    g_\psi  (n)/ g_{\xi_r}(n) < + \infty
    \}. \label{R^+}
\end{eqnarray}
for all $\{ g_{\xi_r}(n) \} _{r \in A, n \in N}$ satisfying the
three conditions: I. Strict monotonicity for all $r \in A$,  II.
Convexity for all $r \in A$ and $n \in N$, and III.
$\underline{\lim} _{n \rightarrow \infty} (g_{\xi_{r_1}(n)} /
g_{\xi_{r_2}(n)}) > 0$ is equivalent to $r_1 \ge r_2$. The proof
is given by the following:  If $\{ g_{\xi_r}(n) \} _{r \in A, n
\in N}$ satisfies the conditions I, II, and III, the corresponding
set of states $\{ \ket{\xi_r} \} _{r \in A}$ for $\{ g_{\xi_r}(n)
\} _{r \in A, n \in N}$ exists and is totally ordered. Since $A
\in \mathbb{R}$ is assumed to be an interval, the two functions
$R^-(\psi)$ and $R^+(\psi)$ defined by Eqs.~(\ref{rminus}) and
(\ref{rplus}), can be represented by Eqs.~(\ref{R^-}) and
(\ref{R^+}) by using Vidal's Theorem.

%%%%%%%%%%%%%%%%%%%%%%%%%%%%%    Proof    %%%%%%%%%%%%%%%%%%%%%%%%%%%%%%%%%

Next, we show that there exists a pairs of genuine infinite
dimensional states which are incomparable to each other. We prove
that the two monotones $R^-(\ket\psi)$ and $R^+(\psi)$ given by
Eqs.~(\ref{R^-}) and (\ref{R^+}) do not necessarily coincide with
each other for infinite dimensional systems, by constructing an
example. We consider a twice continuously differentiable function
$g(x)$, which is the continuous counterpart of $g(n)$, since a
continuous function is more convenient for analytical
investigation.  The conditions for $g(x)$ to relate to a genuine
infinite dimensional state is now given by $g(x)>0$ (strict
positivity), $g^{'}(x) <0$ (strict monotonicity), $g^{''} (x) \ge
0$ (convexity), and $g(0) =1$ (normalization), for all $x$. If
$g(x)$ satisfies the above conditions except the normalization
condition, we can easily normalize $g(x)$.  Thus we omit the
normalization condition for simplicity.  Since convertibility is
determined only by the ratio of functions, we introduce another
function $d(x)$, which is given by $d(x)=p(x)g(x)$.  Let $g(x)$
satisfy the same conditions as $g(x)$. Then $p(x)> 0$, $g^{'}(x)
p(x) + g(x)p^{'} (x)<0$, $g^{''} (x) p(x) + 2g^{'}(x)p^{'}(x)+g(x)
p^{''}(x) \ge 0$ and $p(1) = 1$ are to be satisfied.

We set our function to be $g(x)= e^{-x}$.  Actually, our choice of
$g(x)$ represents one of the most tractable genuine infinite
dimensional entangled states, the two mode squeezed state
$\ket{\psi_q} =\frac{1}{c_q} \sum _{n=0}^{\infty} q^n \ket{n}
\otimes \ket{n}$, where $q$ is a squeezing parameter. We give a
construction of a function $d(x)$ which indicates the existence of
incomparable genuine infinite dimensional states. In this case,
the conditions for $p(x)$ become simple, $p(x) > 0 $, $p(x) -
p^{'}(x)>0$, and $p(x) -2p^{'}(x) +p^{''}(x) \ge0$. We choose
$p(x)$ to be parameterized by $r$ as $p(x)=p_{r}(x) = (\log
x)^{r}\{ \sin (\log x)+1\} +({\log x})^{-1}$ where $0 < r <
+\infty$. We define two functions $m_r(x) \equiv p_r(x) -
p_r^{'}(x)$ and $c_r(x)\equiv p_r(x) -2p_r^{'}(x) +p_r^{''} (x)$
for evaluating monotonicity and convexity, respectively.

For all $0 < r_1 < r_2 < \infty$, there exists $x_{r_1, r_2}
> 0$ such that $m_r(x)>0$ and $c_r(x) \ge 0$ for all $x \ge x_{r_1, r_2}$,
and $r \in [r_1, r_2]$.  That is, the function $p_r(x+x_{r_1,
r_2})$ satisfies the positivity, monotonicity and convexity
conditions. Therefore we can consider a state $\ket{\xi_r}$
represented by the function $d_r(x)=p_r(x+x_{r_1, r_2}) g(x)$. The
ratio of the functions $d_r(x)/g(x)=p_r(x+x_{r_1, r_2})$
determines convertibility between the two states $\ket{\psi_q}$
and $\ket{\xi_r}$. To evaluate the ratio, we rewrite
$p_r(x+x_{r_1, r_2})$ in the discrete form: $p_r(n')=p_r(\Delta n+
x_{r_1, r_2})$ where $\Delta=-\log q$.  Then we can easily show
that $\underline{\lim} _{n \rightarrow \infty} p_r (n) = 0$ and
$\overline{\lim} _{n \rightarrow \infty} p_r (n) = \infty$.
Defining $R^-$ and $R^+$ from $\{ \ket{\xi_r} \}_{r \in
(r_1,r_2)}$, we obtain $R^-(\psi) =  r_1$ and $R^+(\psi) = r_2$.
The two states $\ket{\psi}$ and $\ket{\xi_r}$ for all $r \in
[r_1,r_2]$ are now shown to be incomparable under SLOCC.

%%%%%%%%%%%%%%%%%%%%%infinite number of incomparable class%%%%%%%%%%%%%%%

Further, we show that there are {\it infinitely many} mutually
incomparable states by proving that if $\{ \ket{\Psi _k} \}
^{\infty}_{k=0}$ is defined by the following sequence $\{ d_k
(\Delta n) \} ^{\infty }_{k=0}$ and $d_k(x) = C_k e^{-x} \{p
(x+a_k) \}^k$, where $C_k$ is a normalization constant preserving
$d_k(0) =1$ and $a_k >0$ is a real number, then $\ket{\Psi _k}$ is
SLOCC incomparable each other for all $k \in N$.  It is easy to
prove that for all $k \in N$, there exists a real number $a_k$
such that for all $\{ d_k (\Delta n) \} ^{\infty }_{k=0}$
satisfies the conditions for Vidal's monotone (strict positivity,
strict monotonicity, convexity and normalization) for all $x \ge
a_k$.

To prove the mutual incomparability of $\ket{\Psi _k}$ for
different $k$s, it is sufficient to show $\overline{\lim} _{n
\rightarrow \infty } d_k (\Delta n ) / d_l(\Delta n) = + \infty$
and $\underline{\lim} _{n \rightarrow \infty } d_k (\Delta n ) /
d_l(\Delta n) = 0$ for all $k > l$. The first condition is
evaluated by
\begin{eqnarray*}
    & \quad & \overline{\lim} _{x \rightarrow \infty}
   d_k(x)/d_l(x) \\
    & \ge & \overline{\lim} _{p \rightarrow \infty}
    \frac{ \{ 2 h(p,k)+\frac{1}{h(p,k)} \}^k }
    {[ h(p,l) \{ \sin (h(p,l)) + 1 \}
    + \frac{1}{h(p,l)} ]^l } \\
    & \ge & \overline{\lim} _{p \rightarrow \infty }
    \frac{2^n h(p,k)^n}
    {[ h(p,l) \{ \sin (h(p,l)) + 1 \} + \frac{1}{h(p,l)} ]^l} \\
    & \ge & \overline{\lim} _{p \rightarrow \infty }
    \frac{2^n h(p,k)^n}
    {\{ 2 h(p,l) +\frac{1}{h(p,l)} \}^l} \\
    & = & + \infty,
\end{eqnarray*}
where $h(p,k)=\log(x_p + a_k)$ and $\{ x_p \} _{p=1}^{\infty}$ is
a sequence of positive real values such that $\sin (\log x_p) = 1$
and $x_p < x_{p+1}$.  The second condition is evaluated by
\begin{eqnarray*}
    & \quad & \underline{\lim} _{x \rightarrow \infty}
    d_k(x)/d_l(x)\\
    & \le & \underline{\lim} _{q \rightarrow \infty}
    \frac{{\{ h(q,k) \}^{-k}}}
    { [h(q,l) \{ \sin (h(q,l) +1 \}+
    \frac{1}{h(q,l)}]^l} \\
    & \le & \underline{\lim} _{q \rightarrow \infty}
    \frac{\{ 1/h(q,k)\}^k}{\{ 1/h(q,l)\}^l }\\
    & \le & 0 ,
\end{eqnarray*}
where $\{ x_q \}_{q=1}^{\infty}$ is a positive sequence such that
$\sin (\log x_q) =0$ and $x_q < x_{q+1}$. Therefore, it is proven
that $\ket{\Psi _k}$ is mutually SLOCC incomparable for all $k \in
N$ using Vidal's theorem.

Since $\ket{\Psi_0}$ represent a two mode squeezed state, any
$\ket{\Psi _k}$ with $k>0$ is incomparable to not only the two
mode squeezed state, but also to all $\ket{\Psi _{k^{'}}}$ with
$k^{'} \neq k$.  These family of states can be taken to have all
bounded energy due to the existence of the exponential term
$e^{-x}$ in the Vidal's monotones.  In physically relevant
situations, we should restrict measurement obtaining only finitely
many results for local operations. Shown incomparability is under
stronger local operations including infinitely many measurement
results, the measurement described by the countable infinite
positive operator valued measure. Therefore, the incomparability
property of infinite dimensional states remains under the finite
measurement condition.

We have to note that our functions $R^-(\ket{\psi})$ and
$R^+(\ket{\psi})$ are discontinuous for the usual topology of
Hilbert space due to the discontinuity of the SLOCC convertibility
itself.  However, we can say that the maximum probability to
convert $\ket{\psi '}$ where $\| \ket{\psi} - \ket{\psi '} \| <
\epsilon $ for small $\epsilon $ to $\ket{\phi}$ and the
probability of the inverse process are both very small, if both
$\ket{\psi}$ and $\ket{\phi}$ have infinite Schmidt ranks. The
maximum probability for conversion between the states is
continuous for the topology of Hilbert space.

%%%%%%%%%%%%%%%%%%%%%% summary %%%%%%%%%%%%%%%%%%%%%%%%%%%%%%%

In this letter, we have developed a general formulation for
constructing a pair of convertibility monotones using order
properties.   The monotones are considered as generalizations of
distillable entanglement and entanglement cost. This formulation
can be applied to many different situations to analyze
entanglement convertibility.  We have applied the formulation to
SLOCC convertibility for genuine infinite dimensional pure states
in the single-copy situation.  By constructing an example, we have
proved the existence of infinitely many mutually SLOCC
incomparable pure bipartite states even under the bound energy
condition.

One of the important remarks in this letter is that the ordering
property under SLOCC convertibility is changed radically, from
total ordering to non-total (partial) ordering, with the shift in
dimensionality from finite to infinite.  Even under the
restriction of finite energy and finite measurement conditions,
the infinite dimensionality of Hilbert space offers fundamentally
different entanglement properties from finite dimensional systems.
Although we are only able to treat finite amount of classical
information, the workspace for information processing should not
be considered to be finite for infinite dimensional quantum
systems. Our result encourages the search for other properties of
the ``infinite workspace'', fundamental differences between finite
or infinite dimensional quantum systems, and their use for quantum
information processing.

%%%%%%%%%%%%%%%%%%%%%% acknowledgement %%%%%%%%%%%%%%%%%%%%%%%%%%%%%%%

The authors are grateful to M. Hayashi, M. Ozawa, M.B. Plenio,
W.J. Munro and K. Nemoto for helpful comments. This work is
supported by the Sumitomo Foundation, the Asahi Glass Foundation,
and the Japan Society of the Promotion of Science.

%%%%%%%%%%%%%%%%%%%%%%%%%%bibliography%%%%%%%%%%%%%%%%%%%%%%%%%%%%%%%%%

\end{document}